\def\be{\begin{equation}}
\def\te{\end{equation}}
\def\bea{\begin{eqnarray}}
\def\nn{\nonumber}
\def\tea{\end{eqnarray}}

\def\ha{{1\over 2}}

\def\pa{\left(\xi + \alpha - \frac{1}{6}\right)}

\def\a{\alpha}
\def\b{\beta}

\def\d{\delta}

\def\g{\raisebox{.4ex}{$\gamma$}}

\def\k{\kappa}
\def\l{\lambda}
\def\m{\mu}
\def\n{\nu}
\def\o{\omega}

\def\t{\tau}

\def\z{\zeta}

\def\G{\Gamma}

\def\L{\Lambda}
\def\O{\Omega}

\def\bb{\bibitem}
\newskip\humongous \humongous=0pt plus 1000pt minus 1000pt

\newif\ifdtup

\documentstyle[12pt]{article}
\makeatletter                    
\@addtoreset{equation}{section}  
\makeatother                     

\begin{document}
\title{Relation Between Einstein And Quantum Field Equations}
\author{Leonard Parker\thanks{Electronic address:
leonard@csd.uwm.edu} and Alpan 
Raval\thanks{Electronic address: raval@csd.uwm.edu}\\
{\small Department of Physics, University of Wisconsin-Milwaukee, P.O. Box
413,
 Milwaukee, WI 53201.}\\
{\small WISC-MILW-98-TH-10}}
\date{{\small February 9, 1998}}
\maketitle

\begin{abstract}
\noindent We show that there exists a choice of scalar
field
modes, such
that the evolution of the quantum field in the zero-mass and large-mass
limits is consistent with the 
Einstein equations
for the background geometry. This
choice of modes is also consistent with zero production of these scalar
particles
and thus corresponds to  a preferred vacuum state preserved by the
evolution.
In the
zero-mass limit, we find that the quantum field equation implies 
the 
Einstein equation that determines the scale factor for a
radiation-dominated universe; in the large-mass
case, it implies the corresponding Einstein equation for a
matter-dominated
universe. Conversely, if the classical radiation-dominated or
matter-dominated Einstein equations hold, there is
no production of scalar particles in the zero and large mass limits,
respectively. The suppression of particle production in the large mass
limit is over and above the expected suppression at large mass. Our
results
hold for a certain class of conformally ultrastatic
background geometries and therefore generalize previous results by one
of us for
spatially flat Robertson-Walker background geometries. In these
geometries, we find that the temporal part of the graviton equations
reduces
to the temporal equation for a
massless minimally coupled scalar field, and therefore the results for
massless particle production hold also for gravitons.   
Within the class of modes we study, we also find that the requirement of
zero particle production of massless scalar
particles or gravitons is not consistent with a
non-zero cosmological constant.
Possible implications are discussed.
\end{abstract}

\newpage

\section{Introduction}
It is a well-known prediction of quantum field theory in a
curved background
spacetime that the gravitational field generically
creates particles. This was
first shown in \cite{par1} as an effect for
quantum fields propagating in a spatially flat Robertson-Walker
universe. It was also shown there that the requirement of zero 
particle production by a spatially flat expanding universe
implies the Einstein 
equation, with zero cosmological constant, for the scale factor, in two
limiting cases. In the first case,
the 
requirement of zero production of massless minimally coupled particles,
along with
the assumption that the basic mode functions have their simplest possible 
form, leads
to the Einstein equation for the scale factor of a
radiation-dominated universe. In the
second case, the requirement of zero production of highly massive
minimally coupled 
particles, with the same mode functions, leads to the
Einstein equation for the scale factor of a matter-dominated universe.
Conversely, this means
that there is precisely no creation of massless particles in a
radiation-dominated spatially flat universe, and precisely no creation of
highly massive particles in a matter-dominated (or dust-filled) one.

These results show a consistency between the quantized matter field 
equations and the Einstein equations, though each is formally
independent of the other, and suggest that the full quantized
equations governing matter may, in certain circumstances, imply the
macroscopic Einstein
equations \cite{par1, sakh}.
Based on these results, it was
conjectured that a gravitational Lenz's law
may hold true \cite{par1, Pa}. In other words, the backreaction of
particle
creation and vacuum energy will modify the geometry in such a way as to
reduce the creation
rate, eventually bringing it down to zero, when a self-consistent 
equilibrium-like
state of matter and spacetime is reached in which no further particle
creation occurs. In the low- and high-mass limits, it appears from the
results of Ref. \cite{par1} that such self-consistent states exist in the
context of a quantum field theory of particles.  
For example, if the universe is
saturated with a large number of massless particles with
equation of state $p = (1/3)\rho$, there is no further creation of 
massless particles; and similarly for
the highly massive particles with equation of state $p=0$
\cite{par1, forpar}.
It is
of interest to examine whether the results of Ref. \cite{par1} may be
generalized in the context of a quantum field theory of particles to a
wider class of spacetimes. This paper carries
out such a generalization.   

We would like to mention other related work on similar problems. In
\cite{Pa} it was shown that
the requirement of zero production of massless particles implies the
Einstein equations with zero cosmological constant. We will derive a more
general
version of that result in this paper. The works of Ref. \cite{mot}
suggest other related mechanisms within field theory which 
effectively damp the cosmological constant to zero, after which the
universe enters a Robertson-Walker phase. Also, the requirement of
proportionality
between entropy transfer across a local Rindler horizon and its area has
been shown to
imply the Einstein equations \cite{jacob}.

Let us first summarize briefly the conclusions reached in Ref. 
\cite{par1}.
There, the class of geometries considered are the spatially flat
Robertson-Walker universes, with metric
\be
ds^2 = -dt^2 + a(t)^2(dx^2 + dy^2 + dz^2),
\te
with $a(t)$ being an arbitrary positive function, the scale factor of
the universe.

Consider a minimally coupled scalar field propagating in this
background geometry, with the usual Lagrangian density
\be
{\cal L} = \ha \sqrt{-g}\left(g^{\mu \nu}\nabla_{\mu}\phi
\nabla_{\nu}\phi + m^2 \phi^2\right),
\te
giving rise to the equations of motion
\be
(g^{\mu \nu}\nabla_{\mu}\nabla_{\nu} - m^2)\phi = 0.
\te
We  now separate variables and decompose the field in a box of volume 
$V$, to obtain
\be
\phi ({\bf x}, t) = (2Va^3(t))^{-\ha } \sum_{\bf k}[A_{\bf k}e^{i{\bf
k}\cdot {\bf x}}u_{k}(t)^{\ast} + {\rm H.c.}].
\te
The choice of basis mode functions $u_{k}(t)$ fixes the annihilation
operators $A_{\bf k}$, and hence the vacuum state under consideration.
They satisfy the equation
\be
\label{me}
\ddot{u}_{k} + \left[a^{-2}k^2 - {3\over 4}(a^{-1}\dot{a})^{2}
- {3\over 2}a^{-1}\ddot{a} + m^2\right]u_{k} = 0.
\te
It is useful to define the effective mode frequency $\o _k(t) = (a^{-2}k^2
+ m^2)^{\ha}$, in terms of which the elementary (negative
frequency) solutions to Eq.(\ref{me})
reduce to
$\exp(i\o _k t)$ when $a(t)$ is constant. 

For a general function $a(t)$, no particle creation will
occur for asymptotically static expansions if the exact mode functions are
of the form
\be
\label{adia}
u_k(t) = W_k(t)^{-\ha }\exp\left(i\int^{t}dt'W_k(t')\right),
\te
where the functions $W_k$ are such that they reduce to $\o _k$ whenever
$a(t)$ is constant. We are not requiring that the physical $a(t)$ be
asymptotically static, but are imagining joining two asymptotically static
regions in order to analyze the physical particle content. This is
analogous to adiabatically turning on and off the interaction in
elementary particle scattering theory.

 The mode equation (\ref{me})
with (\ref{adia}) implies that $W_k$ satisfies the following equation:
\be
\label{w1}
W_k^2 - W_k^{\ha }\frac{d^2}{dt^2}W_k^{-\ha} - a^{-2}k^2 + {3\over
4}(a^{-1}\dot{a})^2 + {3\over 2}(a^{-1}\ddot{a}) - m^2 = 0.
\te
The is the equation of evolution for $W_k$. If we now constrain the form
of
$W_k$ such that it is the {\it simplest} form which reduces to $\o _k$
when $a(t)$ is constant, i.e. $W_k(t) = \o _k(t)$, then the above equation
becomes an equation for $\o _k(t)$, and hence a condition on the scale
factor $a(t)$. Let us see how this comes about. 

Substituting $W_k(t) = \o _k(t)$ in (\ref{w1}), we get the equation
\be
\label{geom}
C_1(k,t)(a^{-1}\dot{a})^2 + C_2(k,t)a^{-1}\ddot{a} = 0,
\te
where 
\bea
C_1(k,t) &=& \frac{k^4 + 3m^2k^2 a(t)^2 +(3/4)m^4 a(t)^4}{(k^4 +
m^2a(t)^2)^2},\nn \\
C_2(k,t) &=& \frac{k^2 + (3/2)m^2 a(t)^2}{k^2 + m^2 a(t)^2}.
\tea  
In order that the condition (\ref{geom}) be true for all values of $k$ and
$t$, $C_1$ and $C_2$ must be independent of $k$. This happens only for two
choices of the mass $m$, namely, $m=0$ and in the limit $m\rightarrow
\infty$.

In the massless case ($m=0$), Equation (\ref{geom}) reduces to the
condition
\be
(a^{-1}\dot{a})^2 + a^{-1}\ddot{a} = 0.
\te 
This is recognized as one of the Einstein equations for a radiation
dominated cosmology, with solution $a(t) \sim t^{1/2}$.

On the other hand, in the highly massive case ($m^2 \gg k^2$),
Equation (\ref{geom}) reduces to the condition
\be
{3\over 4}(a^{-1}\dot{a})^2 + {3\over 2}a^{-1}\ddot{a} =0,
\te
which is one of the Einstein equations for a matter dominated (or
dust-filled) universe, with solution $a(t) \sim t^{2/3}$.

The Einstein equation obtained from the zero-particle-creation condition,
in each case, is the one that follows by elimination of the energy density
and pressure from the full set of Einstein equations using the equation of
state. We will call this the purely geometric Einstein equation. Note that
the equations of state also follow from the quantum field equation in the
appropriate limits \cite{parful}.

To summarize, we see that there exists a
certain choice of the function
$W_k$ consistent with zero particle creation, which reduces to the correct
form in flat space, and for which the minimally coupled scalar field
equations imply the
purely geometric Einstein equation in two limits on the mass. 
In the massless case, a
radiation-filled universe is implied, and for the highly massive case, a
dust-filled one. This surprising result  lends support to the
gravitational Lenz's law
conjecture discussed earlier.
The mode
functions corresponding to this choice of $W_k$ therefore define quantum 
states whose particle content is preserved by the evolution, in the zero-
and high-mass limits, when the radiation- and matter-dominated purely
geometric Einstein equation holds,
respectively. We will call such states gravitationally-preferred states.
It seems reasonable to assume that the mode functions in these
gravitationally-preferred states define physical
particles. Here, the Einstein equations replace the timelike symmetry 
used in Minkowski space to define physical particles. We will discuss this 
further in a later
paper \cite{prep}.

In this paper we will show that such 
states exist  more generally, for scalar fields with arbitrary coupling
to curvature, and for gravitons, in
closed and open Robertson-Walker cosmologies, and in a more general
class of conformally ultrastatic spacetimes. In each case we shall show 
that there exists a set of  modes of the form in
Eq.(\ref{adia})(corresponding to zero particle creation), which, when
identified as the exact modes of the scalar field equation, constrain the
geometry to satisfy the purely geometric Einstein field equation. 

The organization of the paper is as follows. In Section $2$, we derive the
flat Robertson-Walker result in a different way, explicitly making 
the connection between the choice of vacuum and the condition on the
geometry, and generalizing this result to the case of
arbitrary coupling to scalar curvature. In Section $3$, we see how the
method of constructing modes in Section $2$ may be used to treat the
closed and open Robertson-Walker (RW) universes. In Section $4$,
we show that this
method can be generalized to treat a certain
class of conformally ultrastatic spacetimes, of which the RW
family is a special case. In Section $5$, we show that the Einstein
equations with non-zero cosmological
constant are inconsistent with zero creation of massless
particles for the general class of geometries considered in Section $4$.
In the case of highly massive particles, this conclusion can be
reached only if the cosmological constant itself does not appear
in the fundamental solutions to the quantum field equations. Finally,
in Section $6$, we show that the graviton equations can be separated  in
this class of geometries, and that their temporal part reduces to
the temporal equation for a massless minimally coupled scalar field, 
thus enabling one to make similar conclusions for gravitons. This
reduction to a minimally coupled scalar field was carried out by Lifshitz
\cite{lifshitz} for gravitons in RW spacetimes, and is here generalized to
background metrics of the form in Eq. (4.10). 
Certain properties of the Einstein equations and a proof of consistency of
the transverse-traceless-harmonic gauge for gravitons are dealt with in
the appendices.

\section{Spatially Flat RW with Curvature Coupling}
In the Introduction, we obtained the Einstein equations for a flat RW
background by making the simplest choice for $W_k$ which was consistent
with zero
particle production for asymptotically static expansion of the universe.
Here we will rederive these  results, generalized to include
arbitrary
curvature coupling of the scalar field, but by asking the question: for
what choice of $W_k$ do the scalar field equations imply the
purely geometric Einstein equation? To answer this, we will introduce a
parametrized 
family of possible
forms of $W_k$ consistent with zero particle production for
asymptotically static expansions and constrain the
value of the parameter such that the Einstein equations hold.
   
We therefore begin by  considering a free quantized scalar field $\phi
(x)$
propagating in a 
spatially flat 
RW spacetime with metric (1.1), with an arbitrary scale
factor 
$a(t)$. The field operator now satisfies 
the Klein-Gordon equation with curvature coupling:
\be
(g^{\mu \nu} \nabla_{\mu} \nabla_{\nu} - m^2 - \xi R) \phi
=0,
\te
The field may be quantized in the usual manner, leading to the mode expansion
\be
\phi({\bf x},t) = (2\pi)^{-3/2} \int d^3k \,(A_{\bf k} e^{i{\bf k}.{\bf
x}}v_k(t) 
+ {\rm H.c.}),
\te
where the continuum limit has been taken in the mode expansion, and
the mode functions $v_k(t)$ differ from the functions $u_k(t)$ in the
Introduction by a factor of $a^{-3/2}$. They  
satisfy the equation
\be
\ddot{v_k}+ 3\frac{\dot{a}}{a}\dot{v_k} + (k^2 a^{-2} + m^2 + \xi R)v_k =
0.
\te
Since RW spacetimes are conformally flat, it is convenient to
work with the 
conformal time $\eta$, 
defined by 
\be
\eta = \int^t dt\, a^{-1}(t),
\te
and also define the conformal factor $C(\eta) = a^2(\eta)$.
With these redefinitions, the metric is given by $ds^2 =
C(\eta)(-d\eta^2 + \sum_i dx^i dx^i)$. 
Furthermore, a redefinition 
of the mode functions,
$\chi_k = C^{1/2}v_k$, eliminates 
the first derivative term in Equation (2.3), and yields 
\be
\label{adia2}
\chi_k'' + \O _k^2(\eta) \chi_k = 0 
\te
where a prime denotes derivative with respect to the conformal time $\eta$.
It is useful to define the effective mode frequency $\O _k$ by
\be
\O _k = \left[k^2 +C\left(m^2 +\left(\xi - 
\frac{1}{6}\right)R\right)\right]^{1/2}. 
\te

Conditions on quantum field behavior are now imposed by demanding a 
particular form  for the solutions to Equation (2.5). We will require 
that these mode functions  are given by
\be
\label{adia3}
\chi_k = (2W_k) ^{-1/2}(\eta)\, \exp \left(i\,\, \int^{\eta} d\eta \,W_k 
(\eta) \right)
\te
where 
\be
\label{w}
W_k = (\O _k^2 + \a C R)^{1/2}.
\te
Note that this definition of $W_k$
differs from the one in the Introduction by a factor of $a^{-1}$. Here,
$\a $ is a dimensionless constant 
parametrizing the family of mode solutions, or equivalently, the family
of vacuum states, to be left arbitrary for now. Also, when $\xi =0$ and 
$\a = 1/6$, 
$W_k = (k^2 + m^2 a^2)^{1/2} = a\o _k$, which was the simple form
discussed in the Introduction. 
We will later find that $\a $ is constrained to take this value for all
$\xi$ by the
requirement that the 
modes satisfy Equation (2.5), and by consistency with the classical Einstein 
equations. The above family of possible solutions to Equation 
(\ref{adia2}) satisfies
the zero particle creation condition, because, when $a(t)$ is constant,
$\chi_k$ becomes a pure negative frequency mode. 

We will now demand that $\chi_k$ satisfy exactly the field equation (2.5). 
We first obtain
\be
\label{e}
\chi_k'' + \O _k^2 \chi_k = \left[ \frac{3}{4}
\frac{W_k'^2}{W_k^2} - \ha \frac{W_k''}{W_k} - W_k^2 + k^2\right. 
\left. \vphantom{\frac{W_k'^2}{W_k^2}}+C\left(m^2 + \left(\xi - {1\over
6}\right)R\right)\right] \chi_k.
\te
The right hand side of the above equation must vanish identically for 
$\chi_k$ to be a mode solution. 
Substituting for $W_k$ from Eq. (\ref{w}), this implies
\bea
\label{ad}
\lefteqn{{5\over 16}W_k^{-4} \left[C'(m^2 + \z R) + \z CR'\right]^2}\nn \\
& & -{1\over 4}
W_k^{-2}\left[C''(m^2 + \z R) + 2\z C'R' + \z C R''\right]- \a CR=0,
\tea
where 
\be
\zeta = \xi +\a -{1\over 6}.
\te
We will use Eq. (\ref{ad}) to obtain a 
condition on the geometry in the two cases when the field is massless, 
and when its mass is very large.\\

\noindent {\bf Zero mass case} \\
\noindent For massless fields,  
Eq. (\ref{ad})  leads to the
condition
\be
{5\over 16}\zeta^2 \left(\frac{d}{d\eta}(CR)\right)^2 - {1\over 4}\zeta
W_k^2 \frac{d^2}{d\eta^2}(CR) - \a CR W_k^4 = 0
\te 
We  wish to fix the constant $\a $ such that the above
equation holds for all $k$. This means that the coefficient of every
power of $k$ must vanish separately. A little algebra then shows that the
only possible solutions are: (i) $\a =0$ and $(\xi - 1/6)CR =$ constant,
or (ii) $\a \neq 0$ and $R=0$. Therefore if we choose some value of $\a
\neq 0$, we must have the condition $R=0$ on the geometry, which is one of
the Einstein equations for a radiation-dominated universe. On the
other hand, if we choose $\a = 0$, there is no condition on the
geometry at conformal coupling $\xi = 1/6$. This happens because the modes
in this case are the exact negative-frequency modes at conformal coupling,
independent of the 
geometry. However, $\a =0$ is certainly consistent with the Einstein
equation $R=0$ for any value of $\xi$, although not a sufficient condition
for it. 

Thus we may conclude from the preceding analysis, that, {\it for any 
value of $\xi$, the choice $\a \neq 0$ implies the classical Einstein
equation determining
the scale
factor for a 
radiation-dominated universe. All choices of $\a $ are consistent with
this equation.} If we assume that $\a $ is independent of the mass, then
the high-mass limit considered below will turn out to exclude $\a =0$. \\

\noindent {\bf Large mass case} \\
\noindent In the limit of mass approaching infinity, 
we may expand
Eq.(\ref{ad}) in
inverse powers
of the mass. The leading term in such an expansion is a term of order
$m^2$ in Eq.(\ref{ad}) whose coefficient vanishes after  substituting for
$W_k$ in that equation. Therefore the leading term in Eq.(\ref{ad}) is of
order 1, and the corrections are of order $m^{-2}$. Keeping order 1 and
order $m^{-2}$ terms in Eq.(\ref{ad}) yields the condition 
\bea
\lefteqn{{5\over 16}C^{-2}C'^2 -{1\over 4}C^{-1}C'' - \a CR
+\frac{1}{4m^2 C}\left[\ha \pa C^{-1}C'R'\right.}\nn \\
& &\left. - \pa R'' +
k^2\left(C^{-2}C'' - {5\over 2}C^{-3}C'^2\right)\right] = 0.
\tea 
If we ignore next to leading corrections in the above equation, we find
that the dominant contribution is a pure condition on the
geometry, independent of mode number. Substituting for $R$ (Eq.(A.13) with
$K=0$), this condition yields
\be
\left({5\over 16} + {3\over 2}\a \right)C^{-2}C'^2 - \left({1\over 4} +
3\a \right)C^{-1}C'' = 0,
\te
which is the correct Einstein equation for a matter-dominated flat RW
universe, if $\a =1/6$. This corresponds to a $W_k$ of the form
\be
\label{W}
W_k^2 = k^2 + C (m^2 + \xi R).
\te
The addition of a $\xi R$ term is related to a shift in the physical
mass \cite{prep}. In the minimally coupled ($\xi =0$) case, this is the
simplest
choice of $W_k$ consistent with zero particle creation. Furthermore, at
minimal coupling, $\zeta$ also vanishes, and the conditions $m^2 \gg \zeta
R$, etc., necessary for the validity of the high-mass expansion of Eq.
(2.10), are automatically satisfied. Therefore there is no condition on
the magnitude of the curvature scalar for the high mass results to hold at
minimal coupling.

 Note that we chose the coefficient of
the $m^2$ term in the definition of $W_k$ (Eq.(2.8)) to be the same as the
coefficient of the $m^2$ term in $\O
_k$. This ensures that the leading term in Eq.(2.10) above is of order 1
rather than order $m^2$. The fact that the coefficient of the $m^2$
term in (2.10) vanishes
means that, as one would expect, there is no particle creation in the
strict $m
\rightarrow \infty$ limit, independent of the geometry. The constraint
(2.13) on the geometry is therefore a next to leading order effect in the
mass.

We thus conclude that {\it for any value of $\xi$, the choice
$\a = 1/6$ in $W_k$, which is consistent with
zero creation of highly massive particles, implies the Einstein 
equation determining the scale factor for a matter-dominated universe.}

Furthermore, if $\a $ is independent of mass, then this 
non-zero value of $\a $ also implies that the
radiation-dominated Einstein equation $R=0$ must hold in the massless
case (for any $\xi$).

We now show that the choice of $W_k$ in the limit of
large mass is unique: that is, we do not need to assume the form 
in Eq.(\ref{w}) for $W_k$ to begin with. In the large mass limit one may
assume the general form $W_k = m (f(\eta) + g(\eta) m^{-2} +$
terms of order $m^{-4})$, where $f(\eta)$ and $g(\eta)$ are unknown
functions. We now substitute this expression for $W_k$ into the
right-hand-side (RHS) of Eq.(\ref{e}), and set it to zero. The resulting
differential equation, to order $1$ in powers of $m^{-2}$, is
\be
m^2(C-f^2)
+\frac{3}{4} \frac{f'^2}{f^2} - \ha \frac{f''}{f} - 2 f \, g +k^2  +
\left(\xi - {1\over 6}\right) CR =0.
\te
Setting the coefficient of $m^2$ to zero yields $f(\eta) = C^{1/2}(\eta)$.
Setting the coefficient of $m^0$ to zero and using the purely
geometric Einstein equation
(A.21) with $K=0$, yields $2g(\eta) = C^{-1/2}(k^2 + \xi CR)$. Therefore,
at large mass, $W_k$ takes the form
\be
W_k = mC^{\ha}\left[1+\ha m^{-2}C^{-1}(k^2 + \xi CR)\right] + {\cal
O}(m^{-3}),
\te
which agrees with a high-mass expansion of $W_k$ as given by Eq.(\ref{W}).
In particular, at minimal coupling, it agrees with the form for the mode
functions given in the Introduction.

\section{Open and Closed RW Spacetimes}

For RW spacetimes with constant spatial curvature, we shall need a slight
modification of the fundamental modes $\chi_k$ in order to generate the
Einstein equations in the two limiting cases of high and zero mass.

The spacetime metric is now given by
\be
ds^2 = C(\eta)(-d\eta^2 + p_{ij}dx^i dx^j),
\te
where
\be
p_{ij}dx^i dx^j = (1-Kr^2)^{-1}dr^2 + r^2 d\O ^2,
\te
with $K=\pm 1$ and 0 for the closed, open
and flat spatial geometries. We  use Latin indices to
denote spatial components. 
Defining $p \equiv {\rm det}(p_{ij})$, the scalar
field equation with curvature term becomes
\be
\label{F}
C^{-1}\partial_{\eta}(C\partial_{\eta})\phi - p^{-\ha}\sum_{i,j}\partial_i
(p^{\ha}p^{ij}\partial_j)\phi + C(m^2 + \xi R)\phi = 0.
\te
This may be solved by separation of variables. Therefore we expand the
scalar
field in terms of the modes
\be
\label{E}
\phi({\bf x}, t) = (2\pi)^{-3/2}\int d\mu(k) (A_{\bf k}\, Y_k({\bf x})
v_k(\eta)
+ {\rm H.c.}),
\te
where the spatial modes $Y_k$ are eigenfunctions of the Laplacian on the
three-space
\cite{parful}. The definition  of the measure $\mu(k)$ for the open,
closed and flat cases may also be found in the same reference.  

The spatial modes satisfy the equation
\be
\label{S}
\Delta^{(3)} Y_k \equiv p^{-\ha}\partial_i(p^{\ha} p^{ij} \partial_j) Y_k
= (K-k^2)Y_k,
\te 
where the quantum numbers $k$ are defined such that $k^2 - K \geq 0$ for
all three values of $K$.

The above equation implies that the temporal modes satisfy
\be
C^{-1}\partial_{\eta}(C\partial_{\eta}) v_k + (C(m^2 + \xi R) + k^2 -
K)v_k = 0.
\te
Introducing the conformal modes $\chi_k = C^{\ha }v_k$ as before, this
equation reduces to the equation
\be
\label{corw}
\chi_k'' + \left[k^2 + \left(m^2 + \left(\xi - {1\over
6}\right)R\right)C\right]\chi_k = 0.
\te

As in the previous section, we will construct solutions to the above
equation which reduce to exact negative frequency modes whenever $C(\eta)$
is
constant (and thus correspond to zero particle creation), and generate the
Einstein equations at large mass and zero mass. It turns out that the
form in Eq.(\ref{adia3}) is sufficient for this purpose, with a slightly
different
choice for the functions $W_k$. We will now require 
\be
\label{freq}
W_k = \left(k^2 + (m^2 + \xi R)C - \b (m^2)K \right)^{\ha},
\te
where $\b (m^2)$ is some dimensionless function of the mass, and
independent of the conformal factor. We require that $W_k$ be real, i.e. 
$k^2-\b (m^2)K \geq k^2 -K \geq 0$, i.e. $\b (m^2)\leq 1$. The precise
values of $\b$ in the two mass limits of interest will
be determined later, by the requirement of consistency with the Einstein
equations.   For now we will only assume that it varies very slowly at
large mass, so that its derivatives do not contribute to the first few
orders in a large mass expansion, which we shall carry out below. 

Demanding that these modes now satisfy  Equation (\ref{corw}) 
leads to the condition 
\be
\label{field1}
\frac{3}{4} W_k^{-2}W_k'^2 - {1\over 2}W_k^{-1} W_k'' -{1\over
6} C R + \b (m^2)K = 0.
\te

In the massless ($m=0$) case, the above condition implies
\be
{5\over 16} \xi^2 \left(\frac{d}{d\eta}(C R)\right)^2 - {1\over 4}\xi
W_k^2\frac{d^2}{d\eta^2}
(CR) - W_k^4\left({1\over 6}C R - \beta (0)K\right) =
0.
\te
The most general solution to the above equation which is independent of
$k$ is $CR = 6\b (0)K$. For this solution to be the purely
geometric Einstein equation,
$R=0$, for a radiation dominated universe, we must have $\b (0)
=0$. This choice for $\b (0)$ is also motivated by the fact that it must be
zero in the massless conformally coupled case for the modes $\chi_k$ to
reduce to the exact solutions in that limit. Note that when $m=0$, these
results mean that the physical mass remains zero.

To analyze the high mass limit, we may expand (\ref{field1}) in powers of
$m^{-2}$, to get 
\be
\frac{5}{16}C^{-2}C'^2 - {1\over 4}C^{-1}C'' - {1\over 6}C R + \b
(\infty)K + {\cal O}(m^{-2}) = 0.
\te
Substituting for the scalar curvature $R$ (see Eq.(A.13)),
this
yields
\be
{3\over 4}C^{-2}C'^2 - C^{-1}C'' - {4\over 3}(1-\b (\infty))K = 0,
\te 
which is the purely geometric Einstein equation (A.21) for a
matter-dominated RW universe,
if $\b (\infty) = 1/4$.

To summarize, we therefore find that the choice (\ref{freq}) for $W_k$
leads
to the correct geometric Einstein equations in the two limiting cases,
provided $\b
(m^2)$ takes the values $\b (0) =0$, and $\b (\infty) = 1/4$. This
restricts the form of $W_k$ in these two limits. In either limit, it is
true that $k^2 - \b K \geq 0$, therefore ensuring that $W_k$ is real when
$C$ is constant.

\section{Conformally Ultrastatic Spacetimes}

We would now like to generalize the previous results by understanding
them better. The purely geometric Einstein equations considered
so far have generalizations to arbitrary spacetimes. The geometric 
equation for a 
radiation-dominated spacetime continues to be $R=0$. For an arbitary
matter-dominated spacetime, the corresponding equation, obtained by
eliminating the pressure $p$ from the Einstein equations using the
equation of state 
$p=0$, is $R_{\mu \nu}u^{\mu}u^{\nu} = (1/2)R$, where $u^{\mu}$ is the
four-velocity of the fluid elements, and therefore a geodesic tangent
vector. This equation is derived in the Appendix, Eqs.(A.10-12). 

The 
behavior of the field equations
at zero mass seems plausible because the chosen modes do not correspond to 
exact solutions when $\xi = 1/6$ (i.e. conformal modes) unless $R=0$.
Hence, the requirement of zero particle creation
at zero mass constrains the geometry to satisfy $R=0$. However, this sort
of argument does not explain the behavior when the mass is large. In that
limit, there exists a choice of the dimensionless parameters in the mode 
functions for all three types of RW
geometries, such that the scalar field equations imply the purely
geometric Einstein equation $R_{\mu
\nu}u^{\mu}u^{\nu} =(1/2) R$. 
The appearance of
$R_{\mu \nu}u^{\mu}u^{\nu}$ in the scalar field equations is rather
mysterious because the scalar field equations do not seem
to depend on timelike geodesics.

However, in Robertson-Walker spacetimes, it is
easy to see that the
geodesic tangent vector field $u \equiv d/dt$ is proportional to the
conformal killing vector
field $d/d\eta$ which generates translations of conformal time.
Directional derivatives along the orbits of this vector
field appear explicitly in the scalar field equations. We thus have a
handle
on how to obtain a generalization of the Robertson-Walker results. We must
 look for spacetimes with a timelike conformal killing vector field,
and require that the integral curves of this vector field are geodesics.

We  therefore consider a general spacetime with an everywhere 
timelike vector
field $b^{\mu}$, and pick the time coordinate $\eta$ to be the affine
parameter along integral curves of $b$. With this choice, we have 
$b^{\mu} = (1, 0, 0, 0)$. We now demand that $b$ be a conformal
killing vector field i.e. the spacetime is conformal to a stationary
spacetime. We will call such a spacetime ``conformally stationary''. This
property implies
\be
\label{kill}
2\nabla_{(\mu}b_{\nu )} = \lambda(x) g_{\mu \nu},
\te
where $\lambda (x)$ is some function. In the chosen
coordinate system, the left hand side of the above equation reduces to
$g_{\mu \nu,0}$. We thus get
\be
g_{\mu \nu,0} = \lambda(x) g_{\mu \nu},
\te
and therefore, in these coordinates, the metric takes the form
\be
g_{\mu \nu}(x) = f_{\mu \nu}({\bf x}) \exp \left(\int^{\eta} d\eta \lambda
(x)\right),
\te
where ${\bf x}$ collectively denotes the spatial coordinates. This is
just a restatement 
of the conformally stationary property.

We further require that $b$ is tangent to a geodesic, and 
therefore satisfies the equation
\be
\label{geo}
b^{\mu}\nabla_{\mu}b_{\nu} = \t (x)b_{\nu}.
\te
Although a rescaling of $b$ by a scalar quantity would lead to an affinely
parametrized geodesic equation, such a rescaling would change the form of
Eq. (\ref{kill}). We find it convenient to keep the form of Eq.
(\ref{kill}) unchanged and allow for a non-affine parametrization in Eq.
(\ref{geo}) above.
Applying $b^{\mu}$ to Equation (\ref{kill}), and using Eq.
(\ref{geo}), we obtain 
\be
\ha \partial_{\nu}(b^2) =  (\lambda(x) - \t (x))b_{\nu}.
\te
In the chosen coordinate system, $b_{\nu} = g_{\nu 0}$. Therefore $b^2
=g_{\mu \nu}b^{\mu}b^{\nu} = g_{00}$, and we get the equation
\be
\ha \partial_{\nu}(g_{00}) = (\lambda(x) - \t (x))g_{\nu 0}.
\te
Using Eq. (4.3), we find that the time component of Eq. (4.6) yields 
\be
\label{con}
\t (x) = \ha \l (x),
\te
and the spatial components then yield
\be
\label{con2}
\l (x) f_{i0}({\bf x}) = \partial_i f_{00}({\bf x}) + f_{00}({\bf
x})\,\partial_i\, \int^{\eta} d\eta \l (x),
\te
where we have used (\ref{con}) in deriving the above equation.

We will now consider a subclass of the metrics which satisfy equations
(\ref{con}) and (\ref{con2}) above. A  restriction which will render the 
field equation separable but is still more general than the RW family of
spacetimes,  is to require
that the vector field $b$ is orthogonal to the hypersurfaces of constant
conformal time $\eta$. Then the spacetime
is conformally static, and the metric components $f_{i0}$ can be chosen to
vanish.
With this staticity assumption, Eq. (\ref{con2}) implies
\be
\exp\left(\int^{\eta} d\eta \l (x) \right) = f_{00}^{-1}({\bf x})C(\eta),
\te
where $C(\eta)$ is an arbitrary function of the conformal time. The metric
may therefore be written as
\be
\label{ultra}
ds^2 = C(\eta)[-d\eta^2 + p_{ij}({\bf x}) dx^i dx^j],
\te
where $p_{ij} = f_{00}^{-1}f_{ij}$ are now arbitrary functions of the
spatial coordinates. The Robertson-Walker family of metrics are the 
special
cases corresponding to  three-spaces
of constant curvature.

We have therefore shown that the most general conformally static metrics
whose conformal killing vector field is tangent to a geodesic, are the
class of conformally ultrastatic metrics given by (\ref{ultra}). 

Our next step is to consider the scalar field equation in the general
spacetime given by (\ref{ultra}). 
We first note that
$CR$ is a sum of function of time and a
function of the spatial coordinates (see Eq.(A.7)). This
separability, combined with the
fact that the metric components $p_{ij}$ are functions of the spatial
coordinates alone, implies that the scalar field equation (\ref{F}) is
separable. The field therefore admits a mode expansion of the form
(\ref{E}), with the spatial modes $Y_k$ satisfying the equation
\be
\left(\Delta^{(3)} + \xi \overline{R}\right)Y_k = -E_k Y_k,
\te
where $\overline{R}$ is the scalar curvature of the ultrastatic metric
conformally related to (\ref{ultra}) (see Eq. (A.8)). 
The  modes $\chi_k(\eta) = C^{\ha}(\eta)v_k(\eta)$ satisfy
the equation
\be
\label{fi}
\chi_k'' + \left[E_k + Cm^2  + \left(\xi - {1\over
6}\right)(CR-\overline{R})\right]\chi_k =0,
\te
where $E_k$ is a separation constant. Note that the quantity $CR -
\overline{R}$ in the above equation is a
function of time alone.

Going through the argument which is by now familiar, we demand
solutions to Equation (\ref{fi}) of the form (\ref{adia3}), with
\be
W_k = (E_k + Cm^2 + \a RC - \b (m^2)\overline{R})^{\ha},
\te
where $\a $ and $\b (m^2)$ are arbitrary quantities to be determined by
consistency with the Einstein equations. This choice of $W_k$ seems
inconsistent with separability of the field equation, because $W_k$ is
clearly not a function of time alone unless either $\b = \a $ or
$\overline{R}$ is a constant. However, we will find that, in the limits 
when the adiabatic modes given by (\ref{adia3}) are exact (massless and 
high mass limits), consistency with the Einstein equations shall force
$\overline{R}$ to be a constant, thus also ensuring that $W_k$, and 
hence $\chi_k$, is a function of time alone. 

The field equation
(\ref{fi}) then implies an equation similar to (\ref{field1}), namely
\be
\label{field2}
{3\over 4}W_k^{-2}W_k'^2 - \ha W_k^{-1}W_k'' - \left(\a + {1\over
6}-\xi\right)CR + \left(\b(m^2) + {1\over 6} - \xi \right)\overline{R} =
0.
\te   
Again, in the massless case, this implies 
\bea
\lefteqn{\frac{5}{16}\a ^2\left(\frac{d}{d\eta}(CR)\right)^2 -{1\over 4} 
\a
W_k^2\frac{d^2}{d\eta^2}(CR)}\nn \\
& & -
\left(\left(\a -
\xi +{1\over 6}\right)CR - \left(\b (0) - \xi +{1\over
6}\right)\overline{R}\right)W_k^4 = 0.
\tea
The most general solution of the above equation which is independent of
$E_k$ is 
\be
\label{sol}
\left({1\over 6} - \xi -\b (0)\right)\overline{R} = \left({1\over 6}-\xi 
 +\a \right)RC,
\te
with the additional condition $RC=$ constant if $\mu \neq 0$. 

Therefore, we must either choose $\b (0) = \xi -1/6$ or have
$\overline{R} = 0$ for consistency with
the Einstein equation $R=0$. Analogous to  the flat RW case
treated in
Section 2, any choice of $\a $ is consistent with this Einstein equation.
More precisely, Eq.(\ref{sol}) {\it implies} the purely geometric Einstein
equation for
all values of $\xi \neq \a + 1/6$. 
We shall now find that  the value of $\a $ required
for consistency with the matter-dominated Einstein equation in the large
mass limit is $\a =\xi$. Therefore, if $\a $ is independent of the mass,
this value also implies the radiation-dominated Einstein equation $R=0$ at
zero mass.
 
In the high mass limit, Eq.(\ref{field2}) implies 
\be
{5\over 16}C^{-2}C'^2 - {1\over 4}C^{-1}C'' + CR\left(\xi -
{1\over 6} - \a \right) + \overline{R}\left(-\xi +{1\over 6} + \b
(\infty)\right) +{\cal O}(m^{-2}) = 0.
\te
Using Equations (A.7) and (A.9)  to express time
derivatives of $C$ as
linear combinations of $CR$, $\overline{R}$ and $R_{00}$, we get
\be
\label{ein1}
{1\over 4}R_{00} + \left(\xi - \a -{1\over 8}\right)CR +\left(-\xi + \b
(\infty) +{1\over 8}\right)\overline{R}  = 0.
\te
We still have freedom to choose $\a $ and $\b (\infty)$. We will choose 
\be
\a = \xi, ~~~ \b (\infty) = \xi - {1\over 8},
\te
so that Equation(\ref{ein1}) becomes  the purely geometric Einstein
equation
\be
R_{00} - \ha RC = 0.
\te
Note that this is the only choice of parameters leading to the above
Einstein equation.

With the choices $\xi -1/6$ and $\xi -1/8$ for $\beta$ in the two limits,
$W_k = (E_k + m^2C + \xi
CR + (1/8 -
\xi)\overline{R})^{1/2}$ at large mass, and $W_k = (E_k + \xi CR + (1/6 -
\xi)\overline{R})^{1/2}$ at zero mass. 

We now show that, in both these
limits,
the relevant Einstein equation forces $\overline{R}$ to be constant, thus
ensuring consistency with separation of the field equation, as mentioned
earlier.

In the massless case, $R=0$ implies
\be
{1\over 3}\overline{R} = \ha C^{-2}C'^2 - C^{-1}C'',
\te
according to Eq.(A.7). Since the left-hand-side of the
above
equation is a function of spatial
coordinate alone, and the right-hand-side a function of time alone, each
quantity must be constant, thus $\overline{R}$ is constant. 

In the high mass case, $2R_{00} = CR$ implies
\be
{1\over 3}\overline{R} = {3\over 2}C^{-2}C'^2 - 2C^{-1}C'',
\te
where we have used Eqs.(A.7) and (A.9).
Again, both sides of the above equation must be separately constant, thus
$\overline{R}$ is constant.

The constancy of $\overline{R}$ in both mass limits implies that  the
conformal factor $C(\eta)$, in both limits, obeys equations identical to
the equations for the conformal factor in the RW
universes. Therefore the function $C(\eta)$ in the general conformally
ultrastatic class is identical to the corresponding function in the RW
universes, and this function will depend on the sign of $\overline{R}$.   
Nevertheless, the constancy of $\overline{R}$ does not necessarily imply 
a RW spacetime.

\section{Inclusion of a Cosmological Constant}

We now consider the question of whether the requirement of zero particle
creation can be made consistent with the Einstein equations with a
non-zero cosmological constant $\Lambda$. In \cite{Pa}, it was shown that
this
cannot be done for massless scalars in Robertson-Walker spacetimes. Here,
we verify that
result using a more general set of allowed modes, and for the
general conformally ultrastatic class of spacetimes. On the other hand,
we will also show that
zero creation of highly massive particles can be made consistent with
the Einstein equations with dust and a non-zero cosmological constant.
However, this requires introduction of $\L $ itself into the form of $W_k$
appearing in the mode functions of Eq. (2.7).
 
To this end, consider Equation (\ref{fi}) for the mode functions in the
metric (\ref{ultra}). We will now demand that this equation be satisfied
by modes of the form (2.7), with an even more general form of $W_k$
than we have considered so far, with more arbitrary parameters. We will
allow $W_k$ to have the form
\be
\label{cosm}
W_k = \left(E_k + C(m^2 +\g \L ) + \m RC - \b (m^2)
\overline{R}\right)^{\ha }.
\te
This choice involves introducing the dimensionful cosmological constant
$\L $, itself, into the fundamental solutions of the field equations, even
though $\L $ does not appear in those equations. This is a rather
unnatural generalization of the form of $W_k$. However, in the massless
case, we
shall find that even
this choice does not
generally permit $\L =0$.

In the massless case, this choice of $W_k$ leads to an equation similar to
(4.15), namely
\bea
\lefteqn{\frac{5}{16}\left(\frac{d}{d\eta}(\m CR + \g C\L )\right)^2 -
{1\over 4} W_k^2\frac{d^2}{d\eta^2}(\m CR + \g C\L )}\nn \\
& &-\left(\left(\m - \xi + {1\over 6}\right)CR - \left(\b (0) - \xi
+{1\over 6}\right)\overline{R} + \g C \L\right)W_k^4 =0.
\tea 
The most general solution of the above equation which is independent of
$E_k$ is given by the pair of equations
\bea
\label{sol1}
\left(\m -\xi +{1\over 6}\right)CR + \g C\L &=& \left(\b (0) - \xi +
{1\over 6}\right)\overline{R} \\
\label{sol2}
C\left(\m R + \g \L \right) &=& \kappa({\bf x})
\tea
where $\kappa$ is an integration constant. For Eq. (\ref{sol2}) to have
the solution $R=4\L $, which is the relevant radiation-dominated Einstein
equation with a cosmological constant, we must have $\kappa =0$ and $\g =
-4 \m $. Then Eq. (\ref{sol1}) becomes
\be
\label{solp}
4\left({1\over 6} - \xi\right)C\L = \left((\b (0) - \xi + {1\over
6}\right)\overline{R}.
\te
Since the left-hand-side of the above equation is a function of time in
general, and the right hand side a function of spatial coordinates, the
only possibilities are:\\
(i) Both $\overline{R}$ and $C$ are constant, which implies a static
universe. \\
(ii) $\b (0)\overline{R} = 0$, $\xi = 1/6$. At conformal coupling,
the field equation (\ref{fi}) for $m=0$ can be explicitly solved and
implies
zero particle  creation for any value of $R$. Thus the conformally coupled
case is trivially consistent with $R=4\L $.\\
(iii) $(\b (0) +1/6 -\xi)\overline{R} =0$ and $\L =0$.\\

\noindent Therefore, if we allow only dynamical spacetimes and
non-conformally coupled fields, the only possibility consistent with zero
creation of massless particles is evidently $\L =0$. This is a more
general
restatement of the results of Ref. \cite{Pa}. We shall further show in the
next section that gravitational perturbation of the class of conformally
ultrastatic spacetimes considered here obey equations of the same form as
the equation for massless minimally coupled scalar fields. Zero creation
of these gravitons must then imply $\L =0$ in a dynamical universe.

We will now go on to show that a non-zero value of $\L $ can be made
consistent with zero creation of highly massive particles for a non-zero
value of the parameter $\g $. In the high
mass limit (we now have the additional requirement that $m^2 \gg \L $),
the field equation (\ref{fi}) with $W_k$ given by (\ref{cosm}) implies an
equation similar to (4.18):
\be
CR\left(\xi -\m -{1\over 8}\right) -\g C\L + {1\over 4}R_{00}
+\overline{R}\left(\b (\infty) - \xi +{1\over 8}\right) =0.
\te
For the above equation to reduce to the matter-dominated Einstein equation
(A.15), we must then have
\bea
\b (\infty) &=& \xi -{1\over 8},\nn \\
\m &=& \xi,\nn \\
\g &=& -{3\over 4},
\tea 
which leads to the form 
\be
W_k = \left(E_k + C\left(m^2 + \xi R - {3\over 4}\L \right) - \left(\xi -
{1\over 8}\right) \overline{R}\right)^{\ha }
\te
at high mass.

Also, an argument similar to the case of zero cosmological
constant shows that $\overline{R}$ must be constant for the
matter-dominated Einstein equations with cosmological constant to hold.
Thus $W_k$ is a function of time alone when the Einstein equations hold,
consistent with separation of the field equation.

\section{Graviton Equations}

In this section, we will show that the temporal part of the graviton
equations in the
conformally ultrastatic metrics (\ref{ultra}) are of the same form, in
Transverse-Traceless-Harmonic (TTH) gauge, as
the temporal equations for massless minimally coupled scalar fields. The
RW case,
for which this is known to be true \cite{lifshitz, fordpar}, is
a special case of this analysis. 

We begin by expressing the Einstein equations in the form
\bea
R_{\m \n } &=& \k \left( T_{\m \n } - \ha g_{\m \n }T \right)\\
\label{ee}
&=& \k \left( u_{\m }u_{\n }(\rho +p) + \ha g_{\m \n }(\rho -p) \right),
\tea
where the second equality holds for a perfect fluid energy-momentum
tensor, given by (A.2). We have also defined $\k = 8\pi G$.

Consider metric perturbations $\d g_{\m \n } = h_{\m \n }$ such that $\d p
= \d \rho = \d u^{\mu} =0$. Unit normalization of the four-velocity
$u^{\mu}$ then implies $h_{\m \n }u^{\m }u^{\n }=0$. For such
perturbations, we may perturb the right-hand-side (RHS) of
Eq.(\ref{ee}) to first order in $h_{\m \n }$ to
write 
\be
\label{tert}
\d R_{\m \n } = \k (\rho +p)u^{\a }u^{\b }(g_{\n \b }h_{\m \a }+g_{\m \a 
}h_{\n \b }) + {1\over 2}\k (\rho -p)h_{\m \n },
\te 
where $g_{\m \n }$ appearing on the RHS is the zeroth order background
metric. For the rest of this section, all covariant derivatives and
curvature tensors will refer to this background metric, which will also be
used for raising and lowering indices.

Furthermore, the perturbed Ricci tensor, to first order in $h_{\m \n }$ is
given by
\bea
\label{rich}
\d R_{\m \n } &=& \ha (2{{h_{(\m }}^{\a }}_{;\n )\a } - h_{\m \n } -
{h_{\m \n ;\a }}^{\a } )\nn \\
&=& \ha ({{h_{\m }}^{\a }}_{;\a \n } + {{h_{\n }}^{\a }}_{;\a \m }+
2R_{\b (\m }{h_{\n )}}^{\b } - 2 {R^{\b }}_{\m \a \n }{h_{\b }}^{\a }-
h_{\m \n } - {h_{\m \n ;\a }}^{\a } ) 
\tea
where $h = g^{\m \n }h_{\m \n }$, and we have used the Ricci identity in
writing the second equality.
At this point, it is useful to introduce the harmonic gauge conditions
\be
\label{har}
{{h_{\m }}^{\a }}_{;\a } =0.
\te
Furthermore, we show in Appendix B that it is consistent to demand the
transverse traceless gauge conditions $h_{\m \n }u^{\n }= h =0$ for the
conformally ultrastatic class of background metrics.
Combining these conditions with Eqs. (6.3-5), we therefore get
\be
\label{ge}
{h_{\m \n ;\a }}^{\a } - 2R_{\b (\m }{h_{\n )}}^{\b } + 2{R^{\b }}_{\m \a
\n }{h_{\b }}^{\a } = \k (\rho -p)h_{\m \n }.
\te
We can now use the zeroth order perfect fluid Einstein equations to
express $\rho$ and $p$ in terms of the background curvature. This yields
$\k (\rho -p) =
(2/3)(R+R_{\m \n }u^{\m }u^{\n })$, and Eq. (\ref{ge}) finally takes the
form
\be
{h_{\m \n ;\a }}^{\a } -2R_{\b (\m }{h_{\n )}}^{\b } +2{R^{\b }}_{\m \a \n
}{h_{\b }}^{\a } -{2\over 3}(R+R_{\a \b }u^{\a }u^{\b })h_{\m \n } = 0.
\te 
We may use Eqs. (A.4-7) to explicitly evaluate the various terms in the
above
equation, and note that $h_{0\m } =0$ because of the transverse gauge.
Then we get
\be
\label{gra}
h_{m n}'' + C^{-1}C'h_{mn}' 
+ {2\over 3}\overline{R} h_{mn} - p^{ij}\left[\overline{\nabla}_i
\overline{\nabla}_j h_{mn} -
2\overline{R}_{i(n}h_{m)j} + 2 {\overline{R}^l}_{min}h_{lj}\right]=0, 
\te
where all barred quantities are evaluated in the ultrastatic metric (A.8).
Recall that prime refers to partial derivative with respect to $\eta$. We
now separate variables in Eq.(\ref{gra}) by writing $h_{mn} = \psi
(\eta) H_{mn}({\bf x})$. This yields the equations
\be
p^{ij}\left[\overline{\nabla}_i
\overline{\nabla}_j H_{mn} -
2\overline{R}_{i(n}H_{m)j} + 2 {\overline{R}^l}_{min}H_{lj}\right]
+\left(E_k+ {2\over 3}\overline{R}\right) H_{mn}=0,
\te  
for the spatial part, and
\be
\label{mincop}
\psi'' + C^{-1}C'\psi' +E_k\psi = 0
\te
for the temporal part. Here, $E_k$ is a separation constant.
In separated variables, the harmonic gauge condition can be expressed as 
\be
\overline{\nabla}_j{H^j}_i =0,
\te
and the traceless condition as
\be
{H_i}^i =0.
\te
The gauge conditions therefore involve only  the spatial modes. The
equation (\ref{mincop}) for the temporal modes may be rewritten by
defining the conformal modes $X(\eta) = C^{\ha}(\eta)\psi(\eta)$. Using
Eq.(\ref{mincop}), these modes
can
be shown to obey the
equation 
\be
\label{min}
X'' + \left[E_k -{1\over 6}(CR-\overline{R})\right]X=0.
\te
Comparing this with Eq.(4.12) for the scalar field modes, we have
therefore found that the temporal modes $X(\eta)$ satisfy the same
equation as temporal modes of a minimally coupled massless scalar field.
Since the analysis of massless scalar field modes in Sections $4$
and $5$ relied
only on the temporal equation, the conclusions of that analysis also apply
for gravitons. Specifically, the condition of zero graviton creation is
consistent with the radiation-dominated Einstein equation {\it without } a
cosmological constant.

\section{Summary and Conclusions}

To summarize our results, we have considered spatially flat
RW spacetimes, spatially curved RW spacetimes, and finally a class of
conformally ultrastatic spacetimes. For each case, we have shown that
there exists a family of functions with the
following properties:\\
(i) They reduce to pure negative frequency temporal mode solutions of the
scalar field and graviton equations during any period when
$C(\eta)$ is constant. [In particular, this holds whenever the first and
second time derivatives of $C$ vanish.]\\
(ii) When $C(\eta )$ is not constant, they are exact mode solutions of
the scalar field and graviton equations in the zero mass limit, and
the scalar field equation in the high mass limit,
if the radiation-dominated and matter-dominated Einstein equations hold,
respectively; and conversely, if these mode functions are exact then 
the Einstein equation that determines $C(\eta)$
must hold in each mass limit.

These properties imply that there is no mixing of positive and negative
frequencies (i.e. no particle creation) between any two periods with
constant $C(\eta)$.  
For the RW family, this means that 
the condition
of zero particle creation is consistent with the Einstein equations in the
limits of zero mass and large mass, and the mode functions we have found
give rise to the gravitationally-preferred states defined in the
Introduction. Recall that property (i) does not require that there be
actual periods in which $C(\eta)$ is static or slowly varying. Rather,
property (i) is similar to the mathematical device of adiabatically
turning on and off the interaction in analyzing the scattering of
elementary particles. For the more general class of conformally
ultrastatic spacetimes given by Eq. (4.10) with arbitrary three-metric 
$h_{ij}({\bf x})$,
it is still plausible that when the chosen temporal modes are exact, 
there is no particle creation due to the spatial
dependence of the metric.
Firstly, it does not seem possible to incorporate an event horizon for the
class of conformally ultrastatic metrics 
considered because the black hole metrics 
cannot be expressed in the conformally ultrastatic form of Eq. (4.10) 
(in which $C$ is only a function of time).
Furthermore, in the absence of a cosmological constant, there are no 
de Sitter - like solutions to the radiation- and matter-dominated equations 
for $C(\eta)$ (such solutions would, in any event, not lead to production of 
real particles, although a monopole particle detector on a geodesic would be
excited by vacuum fluctuations). Also, if
the 3-metric $h_{ij}$ has a scalar curvature singularity (a possible
source of
particle creation) at some point, it must be singular everywhere
because the Einstein equations force $\overline{R}$ to be constant 
everywhere in the spacetime. We cannot, therefore, consider
3-metrics $h_{ij}$ with scalar curvature singularities. In the
absence of horizons and such singularities, it then seems
plausible that there is no particle creation from the spatial
variation of the metric\footnote{We have not proved that there can be no
particle creation due to the spatial variations of the metric. For
example, one could certainly consider three-geometries $h_{ij}({\bf x})$
such that the Riemann tensor or other curvature quantities go singular
somewhere without the curvature scalar being singular. Our conclusions
relating the Einstein equations to zero particle creation would 
not be expected to hold in such situations.}. We then have
gravitationally-preferred states, i.e., consistency
between the Einstein equations
and the condition of zero particle creation in the conformally ultrastatic  
case as well.

In the massless case, one finds that the condition for zero particle creation
for all values of $\xi$ is $R=0$. Note that if we restrict ourselves to
$\xi =1/6$ to begin with, then there is no particle creation even when 
$R \neq 0$. However, the requirement that the gravitational field  
giving rise to zero particle creation vary continously as a function of 
$\xi$ implies that $R=0$ also for $\xi = 1/6$. It is worth pointing out
that 
the conformally ultrastatic metrics are not, in general, conformally flat.
Nevertheless, we find that the mode equation (4.12) at $m=0$ and $\xi =1/6$
(conformal coupling) can be solved exactly and implies zero particle creation
for any value of $R$.
This result is surprising because the usual proof of zero particle creation
at conformal coupling and zero mass makes use of conformal flatness, which
is not present in this case.

Also, it was shown in \cite{Pa} that zero creation of massless scalars or
gravitons in RW universes is not consistent with a non-zero cosmological
constant.
That conclusion was, however, based on a rather restrictive assumption
about the form of the mode functions. The analysis of Section $5$, on the
other hand, comes to the same conclusion by allowing for more parameters
characterizing the modes, and holds
in a more general class of spacetimes. Similarly the graviton analysis of
Section $6$ generalizes the known results for the RW cases to the
conformally ultrastatic spacetimes. 

Interpretation of the results for highly massive scalar fields is not
quite so
straightforward. Here, one does not expect particle production in the
limit of infinite mass anyway, independent of the geometry \cite{par1}.
However, the background geometry will, in general, determine the rate at
which particle production vanishes as we approach infinite mass. Within
the class of background geometries we have considered, we have shown that
the chosen form of $W_k$ differs from the exact form by terms of order
$m^{-2}$
only if the purely geometric Einstein equation is satisfied. If this
Einstein
equation does not hold, then the
chosen form of $W_k$ will differ from the
exact form by terms of order 1 in an expansion in inverse powers of the
mass. Therefore, if the Einstein equations do hold, the particle creation
rate in the high mass limit should converge to zero faster than the
expected rate for an arbitrary geometry. It should be noted that the high 
mass limit actually holds for any value of the curvature scalar at minimal 
coupling, 
because the condition of high mass, 
$m^2 \gg \xi R$, is always true for minimal coupling.

Based on the Lenz law conjecture, we speculate that the results of this
paper would hold in the late time
limit of a dynamical process in which the condition of zero particle
creation would serve as an attractor. This can be tested by setting up a
semiclassical
backreaction computation whose starting point is a choice of initial
geometry, and a quantum state based on a set of scalar field modes in the 
chosen geometry.
After allowing the coupled Einstein and quantum scalar field equations to
evolve,
it should be possible to check if the geometry evolves to one of the
forms given in this paper at late times. Hence, each mode of the scalar
field will evolve to some superposition of the modes constructed in this 
paper; and there will be no further particle production in these modes. In
order for this to
happen, there must be a large amount of initial particle creation by the
geometry,
generating strong backreaction effects, and leading to an effectively
classical energy-momentum tensor at late times.
Therefore, it is necessary that we allow for large curvatures. This is
consistent with
our treatment, for the high mass cases with $\xi =0$, and for the zero mass 
cases with any value of $\xi$. 
For high mass with nonminimal 
coupling, although the scalar curvature is constrained to be small for our
analysis to hold, it is still  possible for other curvature
invariants to be large, leading to a large amount of particle production. 
A concrete backreaction calculation, also addressing questions of
time-reversal invariance, correlations, and admissible initial states will
be carried out in a later paper \cite{prep}.

It is necessary to understand that the gravitational Lenz's law does
{\it not} 
imply that there is no particle creation for any solution of the
Einstein equations, but rather implies that the backreaction, if
sufficiently large, would
drive the system toward  
an equilibrium-like state of matter and geometry. 
A  well-studied example is
that of cosmological anisotropy damping  \cite{ani}. 
Anisotropically evolving universes (such as the Kasner solutions) are
exact solutions of the Einstein equations
in which particle creation can take place. However, the
backreaction of the created particles in the early stages of an
anisotropic universe tends to
drive the anisotropy rapidly to zero, thus inhibiting further particle
creation, and leading to a Robertson-Walker spacetime.
An example in which graviton production in an isotropically
expanding universe gives rise to a radiation-dominated universe
with no further graviton production, is treated in Ref.
\cite{hupa}. Another
case of interest is that of an isolated black hole emitting
Hawking radiation. In this case, however, it does not seem possible to
test the Lenz law conjecture because any equilibrium-like configuration is
expected to occur, if at all, when the hole reaches Planck size, and
semiclassical
gravity theory breaks down.  

Finally, we emphasize that the mode functions in a gravitationally
preferred state (i.e., in one of the quasi-equilibrium states) evidently
give a preferred definition of physical particles even though there is no
timelike Killing vector field. This interpretation has implications to be
discussed in a later work \cite{prep}.\\

\noindent {\bf Acknowledgements}

\noindent This research was supported by the National Science
Foundation under  Grant No. PHY95-07740.

\appendix

\section{Perfect Fluid Einstein equations}

We review the classical Einstein equations with perfect fluid 
matter, in the two cases when the matter is pure radiation or pure dust. 
In the metric signature convention $(-+++)$, the 
Einstein equations take the form
\be
\label{einstein}
R_{\mu \nu} - {1\over 2}g_{\mu \nu}R + \Lambda g_{\mu \nu}= 8\pi G T_{\mu
\nu},
\te 
with the perfect fluid energy momentum 
tensor  given by 
\be
T_{\mu \nu} = pg_{\mu \nu} + (p+\rho)u_{\mu}u_{\nu}, 
\te
where $u^{\mu}$ is the four-velocity of the fluid 
elements, and $p$ and $\rho$ are the principal pressure and density of 
the fluid respectively. A pure radiation fluid satisfies the equation of 
state $p = \rho /3$, and a pure dust fluid satisfies $p=0$. We will
consider these two cases separately for the class of metrics
\be
ds^2 = C(\eta)(-d\eta^2 + p_{ij}({\bf x})dx^i dx^j),
\te
with ${\bf x}$ denoting the spatial coordinates.

\noindent For this class of metrics, we may evaluate the affine
connection and the various curvature tensors, to get
\bea
\label{curv}
{\Gamma^{\a }}_{\m \n }(x) &=& {\overline{\G }^{\a }}_{\m \n }({\bf x}) +
\ha C^{-1}C'\left(2\d ^{\a }_{(\n }\d ^{0}_{\m )} + \d ^{\a }_0
\overline{g}_{\m \n }\right) \\
{R^{\a }}_{\m \b \n }(x) &=& {\overline{R}^{\a }}_{\m \b \n }({\bf x}) +
\ha (\d ^{\a }_{[\n }\d ^0_{\b ]} \d ^0_{\m } + \overline{g}_{\m [\n }\d
^0_{\b ]} \d ^{\a }_0) \left(2C^{-1}C'' - 3C^{-2}C'^2\right)
\nn\\
& &~+ \ha
\d^{\a }_{[\b }\overline{g}_{\n ]\m }C^{-2}C'^2\\
R_{\m \n }(x) &=& \overline{R}_{\m \n }({\bf x}) - \ha \d ^0_{\m } \d
^0_{\n } (2C^{-1}C'' - 3C^{-2}C'^2) + \ha \overline{g}_{\m \n
}\,C^{-1}C''\\
R(x) &=& C^{-1}\overline{R}({\bf x}) + 3C^{-1}\left(C^{-1}C'' - \ha
C^{-2}C'^2\right),
\tea
where all barred quantities are evaluated in the conformally related
metric
\be
d\overline{s}^2 = -d\eta^2 + p_{ij}({\bf x})dx^i dx^j,
\te
and therefore depend only on the spatial coordinates. Also,
note that we have defined $\overline{R}$ as the 4-dimensional scalar
curvature of the metric (A.8) rather than the curvature
associated with the three-metric $p_{ij}$. Furthermore,
the only non-zero components of ${\overline{\G }^{\a }}_{\m \n },
\overline{R}_{\m \n }$ and ${\overline{R}^{\a }}_{\m \b \n }$ are their
spatial components.

\noindent In particular, the time-time component of the Ricci tensor is
given by
\be
R_{00} = {3\over 2}\left(C^{-2}C'^2 - C^{-1}C''\right),
\te
which is independent of the three-metric $p_{ij}$.
\\
\noindent Consider now the Einstein equations (\ref{einstein}), first for
the case
$\Lambda =0$.

\noindent For a radiation-dominated spacetime, the  
equation of state is $p= \rho 
/3$, which yields $T_{\mu}^{\mu} = 0$, and the Einstein equations then
imply $R=0$. \\

\noindent For a matter-dominated (dust) spacetime, the  
equation of state is $p=0$. 
Contraction of Eq.(\ref{einstein})(with $\Lambda=0$) then yields 
\be
R = -8\pi G \rho.
\te
On the other hand, after multiplying Eq.(\ref{einstein}) by
$u^{\mu}u^{\nu}$ and 
summing over repeated indices, we get
\be
R_{\mu \nu} u^{\mu} u^{\nu} + {1\over 2}R = -8\pi G\rho.
\te
The two equations above together imply
\be
\label{dust}
R_{\mu \nu}u^{\mu} u^{\nu} = {1\over 2}R
\te
for any  dust-filled spacetime. For the particular class of metrics we
consider, $u = C^{-1/2}d/d\eta$, and Eq.(\ref{dust}) becomes
\be
R_{00} = {1\over 2}C R.
\te
\\
\noindent If we include the cosmological constant, the analogous equations
are 
\be
R=4\Lambda
\te
for the radiation-dominated case, and 
\be
R_{\mu \nu}u^{\mu}u^{\nu} = \ha R - 3\Lambda
\te
for the matter-dominated case.\\

\noindent {\bf Robertson-Walker Spacetimes}\\

\noindent If we now assume that the spacetime is homogenous and isotropic, 
then it is described by the Robertson-Walker family of metrics 
of the form 
\be
ds^2 = -dt^2 + a^2 (t)\left((1-Kr^2)^{-1}dr^2 + r^2 d\O ^2\right),
\te
where $K= \pm 1$ and 0 denoting the spatially closed, open and flat
cases.  Equivalently, in terms of the conformal time, the metric is
\be
ds^2 = C(\eta)\left(-d\eta^2 + (1-Kr^2)^{-1}dr^2 + r^2 d\O ^2 \right),
\te
where $C(\eta) = a^2(\eta)$.
The scalar curvature is
\be
CR = 6K + 3\left(C^{-1}C'' - \ha C^{-2}C'^2\right),
\te
and the time-time component of the Ricci tensor is given by Eq.(A.6) 
above.

We will now consider the Einstein equations without cosmological constant.
The Einstein equation $R=0$ for the conformal factor
of a radiation-dominated universe  then takes the form
\be
C^{-1}C'' - \ha C^{-2}C'^2 = -2K,
\te
or equivalently, in terms of cosmic time and the scale factor,
\be
\dot{a}^2 + a \ddot{a} = -K.
\te
For $K=0$ (the flat case), one may solve the above equations to get $a(t)
\sim t^{1/2}$, or
$C(\eta) \sim \eta^2$.

The Einstein equation $R_{00} = (1/2)CR$ for the conformal factor of a
matter-dominated universe takes the form
\be
2C^{-1}C'' - {3\over 2} C^{-2}C'^2 = -2K,
\te 
or
\be
\dot{a}^2 + 2a \ddot{a} = -K.
\te
Again, for $K=0$, we get $a(t) \sim t^{3/2}$, or $C(\eta) \sim \eta^4$.
\\
\section{Transverse Traceless Condition for Gravitons}
Here, we show that the transverse traceless (TT) condition on the
gravitational perturbations is consistent with
their dynamics, after fixing the harmonic gauge (6.5). This is
certainly true for the RW spacetimes, as shown by Lifshitz
\cite{lifshitz} (see also \cite{fordpar}). We show that it is true for the
general conformally ultrastatic class given by the metric (\ref{ultra}).

We begin with the linearized Einstein equations in harmonic gauge,
obtained by combining Eqs.(6.3-5):
\bea
\label{pert}
\lefteqn{{h_{\mu \nu;\a }}^{\a } + h_{;\m \n } - R_{\b \n }{h_{\m }}^{\b }
- R_{\b \m }{h_{\n }}^{\b } + 2 {R^{\b }}_{\m \a \n }{h_{\b }}^{\a }=}\nn
\\ 
& &~~~ \ha (\rho - p)h_{\m \n } - (\rho +p)u^{\a }u^{\b }(g_{\n \b }h_{\m
\a
} + g_{\m \a }h_{\n \b }).
\tea
The above equation is a second order differential equation for $h_{\mu
\nu}$ and its solution may be specified by specifying $h_{\mu
\nu}\mid_{\eta_0}$ and its time derivative $u^{\a }\nabla_{\a }h_{\m \n
}\,\mid_{\eta_0}$ on some initial spacelike hypersurface $\eta = \eta_0$.
We will choose this initial data such that $h = u^{\a }\nabla_{\a }h =
u^{\m }h_{\m \n } = u^{\a }\nabla_{\a }(u^{\m } h_{\m \n }) =0$
initially, i.e. at $\eta =
\eta_0$. We will now show that these conditions are preserved by the
equations of motion (\ref{pert}) and this will allow us to conclude that
there exist dynamical perturbations of this class of spacetimes which
satisfy $h = u^{\mu}h_{\m \n } = 0$ for all time, i.e. they are traceless
and transverse.

We first consider the traceless condition. Taking the trace of
Eq.(\ref{pert}), we get
\be
\label{tr}
{h_{;\a }}^{\a } + {1\over 4}(\rho - p)h =0,
\te  
where we have used the condition $h_{\m \n }u^{\m }u^{\n }=0$ demanded by
the unit normalization of the four-velocity.
The above equation is an equation for the trace of $h_{\mu \nu}$ which
must be satisfied by every solution of Eq.(\ref{pert}). Its solution is
unique once we specify the initial conditions on $h$ and its time
derivative. The chosen initial conditions $h = u^{\m }\nabla_{\m }h = 0$
thus imply the unique solution $h=0$ of Eq.(\ref{tr}). Therefore
there exist traceless perturbations with $h=0$ for all time. 

For traceless
perturbations, the dynamical equations (\ref{pert}) reduce to the form
\be
\label{pert2}
{h_{\mu \nu;\a }}^{\a } - 2R_{\b (\n }{h_{\m )}}^{\b }
+ 2 {R^{\b }}_{\m \a \n }{h_{\b }}^{\a }=
\ha (\rho - p)h_{\m \n } - (\rho +p)u^{\a }u^{\b }(g_{\n \b }h_{\m \a
} + g_{\m \a }h_{\n \b }).
\te
Showing that the condition $u^{\m }h_{\m \n }=0$ is preserved by the above
equation requires a bit more work. First, we multiply 
by $u^{\m }$, to get
\be
\label{tran}
u^{\m }{h_{\m \n ;\a }}^{\a }- 2u^{\m }R_{\b (\n }{h_{\m )}}^{\b }+ 2
u^{\m }{R^{\b }}_{\m \a \n }{h_{\b }}^{\a } = \ha (3\rho -p).
\te
We wish to simplify this equation in the background metric of
Eq.(\ref{ultra}). In the chosen coordinate system, we then have
$u^{\m } = C^{-1/2}\d ^{\m }_0$. Consider the first term in the above
equation. We may reexpress it in the form
\be
\label{ex}
u^{\m }{h_{\m \n ;\a }}^{\a } = {\left(u^{\m }h_{\m \n }\right)_{;\a}}^{\a
} - {{u^{\m }}_{;\a }}^{\a }h_{\m \n } - 2 {u^{\m }}_{;\a }{h_{\m \n
}}^{;\a
},
\te
and use Eq.(A.4) for the affine connection to evaluate 
\be
{u^{\a }}_{;\b } = \ha C' C^{-3/2} \left(\d _{\b }^{\a } - \d _{\b
}^{0}\d
^{\a }_{0}\right).
\te
After some straightforward but tedious simplifications, one then obtains
\be
{{u^{\m }}_{;\a }}^{\a }= -\ha \d ^{\m }_0 C^{-7/2}C'^2.
\te 
Substituting the two equations above in Eq.(\ref{ex}), and using the
harmonic gauge condition, we get
\be
u^{\m } {h_{\m \n ;\a }}^{\a } = {\left(u^{\m }h_{\m \n }\right)_{;\a
}}^{\a } - C^{-3/2}C'u^{\a }\left(u^{\m }h_{\m \n }\right)_{;\a } -
C^{-3}C'^2 u^{\m }h_{\m \n }.
\te
We have thus expressed the first term in Eq.(\ref{tran}) entirely in terms
of $Y_{\n } \equiv u^{\m }h_{\m \n }$ and its derivatives. Similarly, we
reexpress other
terms, using the following relations which are easily derived using
Eqs.(A.5-7):
\bea
u^{\m }R_{\b \m }{h_{\n }}^{\b } &=& {3\over 2}C^{-2}(C'' -
C^{-1}C'^2)Y_{\n }  \\
u^{\m }{R^{\b }}_{\m \a \n }{h_{\b }}^{\a } &=& -\ha C^{-2}(C'' -
C^{-1}C'^2)Y_{\n }.
\tea  
Equations (\ref{tran}) therefore reduce to differential equations for
$Y_{\n }$:
\be
{Y_{\n ;\a }}^{\a } -C^{-3/2}C' u^{\a }Y_{\n ;\a } + \ha C^{-2}
(3C^{-1}C'^2 - 5C'')Y_{\n } - {R^{\b }}_{\n }Y_{\b } = \ha (3\rho
-p)Y_{\nu}.
\te
Again the initial conditions $Y_{\n } = u^{\a }Y_{\n ;\a }=0$ imply the
unique solution $Y_{\n }=0$ to the above hyperbolic partial
differential equation. This completes the
proof of consistency of the transverse traceless condition with the
equations of motion for the perturbations. 
\\

\end{document}